\documentclass[prb,aps,reprint,showpacs,amsmath,superscriptaddress]{revtex4-1}

\usepackage{bm}
\usepackage{graphicx}
\usepackage{dsfont}

\usepackage{mathrsfs}

\usepackage{framed}
\usepackage{multirow}
\usepackage[all]{xy}

\usepackage{framed}
\usepackage{comment}
\usepackage{here}                    
\usepackage{latexsym}		     
\usepackage{amsmath}     
\usepackage{amsfonts}  
\usepackage{amssymb}
\usepackage{amsthm}
\usepackage{dcolumn}
\usepackage{color}
\usepackage{mathrsfs}
\usepackage{dsfont}

\newtheorem{definition}{Definition}

\newtheorem{theorem}{Theorem}

\renewcommand{\phi}{\varphi}
\renewcommand{\>}{\rangle}
\newcommand{\<}{\langle}

\newcommand{\ket}[1]{|#1\>}
\newcommand{\bra}[1]{\<#1|}

\newcommand{\be}{\begin{equation}}
\newcommand{\ee}{\end{equation}}
\newcommand{\bea}{\begin{eqnarray}}
\newcommand{\eea}{\end{eqnarray}}

\def\unity{\openone}

\def\unity{\mbox{\rm\bf 1}}
\newcommand{\olap}[2]{\langle #1 | #2 \rangle}

\renewcommand{\phi}{\varphi}
\renewcommand{\unity}{\mathds{1}}

\usepackage[colorlinks,citecolor=blue,linkcolor=blue,urlcolor=blue]{hyperref}

\begin{document}

\title{Information-theoretically secure data origin authentication with quantum and classical resources}

\author{Georgios M. Nikolopoulos}

\affiliation{Institute of Electronic Structure and Laser, Foundation for Research and Technology-Hellas, GR-70013 Heraklion, Greece}

\affiliation{Institut f\"ur Angewandte Physik, Technische Universit\"at Darmstadt, D-64289 Darmstadt, Germany}

\author{Marc Fischlin}
\affiliation{Cryptoplexity, Technische Universit\"at Darmstadt, D-64289 Darmstadt, Germany}

\date{\today}

\begin{abstract} 
In conventional cryptography, information-theoretically secure message authentication can be achieved by means of universal hash functions, and requires that the two legitimate users share a random secret key, which is twice as long as the message. We address the question as of whether quantum resources can offer any advantage over classical unconditionally secure message authentication codes. It is shown that passive prepare-and-measure quantum message-authentication schemes cannot do better than their classical counterparts. Subsequently we present an interactive entanglement-assisted scheme,  which ideally allows for the authentication of classical messages with a classical key, which is as long as the message. 
\end{abstract}


\maketitle

	
\section{Introduction}
\label{sec1}

One of the main information security objectives is to provide  assurance about the original source of a received message \cite{book1,book2,book3,book4}.  
This goal is usually referred to as {\em data origin authentication}, 
and it is a stronger version of another cryptographic goal, the so-called {\em data integrity}. The latter addresses the unauthorized (including accidental)  alteration of the message, from the time it was created,  transmitted or stored by an authorized user. 
Data origin authentication, on the other hand, provides assurance of the identity of the sender of the message, in addition to data integrity. 
From another point of view,  data origin authentication implicitly provides data integrity, in the sense that if the message is somehow modified,  then essentially the source of the message has changed. 

Another security objective that is closely related to data origin authentication, is the so-called {\em non-repudiation}, which 
prevents the original sender of a specific message from denying to a third party his/her action. This is a stronger requirement than data origin authentication, because the latter provides this assurance to the receiver of the message, but it does not provide any evidence that could be presented to a third party in order, for example, to resolve a dispute between the sender and the receiver. 

In modern cryptography, data origin authentication is provided by 
{\em message-authentication codes} (MACs), which require the two legitimate users (sender and receiver) to share a common secret key \cite{book1,book2,book3,book4,Abidin2013}. 
Typically, MACs are built from block ciphers \cite{book1,BKR00} or hash functions \cite{book1,book3,book4,Abidin2013,WC81}. 
A MAC takes as input the message and the key,  and produces an authentication tag, which is sent from the sender to the receiver, together with the message. MACs that rely on block ciphers or collision resistant hash functions offer computational security, as opposed to the Wegman and Carter scheme, which exploits universal hash functions and offers information-theoretic (unconditional) security \cite{book3,book4,Abidin2013,WC81}. On the other hand, {\em digital signature schemes} are the main mechanism for providing non-repudiation and they typically rely on public-key cryptography \cite{book1,book2}. 
Both of MACs and digital-signature schemes offer data origin authentication, but only the latter can ensure non-repudiation. Hence, digital-signature schemes are of vital importance in cases where the sender and the receiver of the message do not trust each other (e.g., business applications), and there is the potential of dispute over the message. MACs are widely used in every task, when non-repudiation is not an issue (e.g., in authenticated end-to-end communication), mainly because the computational cost associated with the generation and the verification of a tag is considerably smaller than for digital signatures. It is worth noting, however, that MACs can provide non-repudiation in certain realistic scenarios e.g.,  when a trusted third party is involved \cite{book4}.

Data origin authentication and data integrity have been also 
discussed in a quantum setting. More precisely, Curty and Santos 
have proposed an authentication protocol for 1-bit message, 
which requires the two legitimate users to share a maximally 
entangled state of qubits \cite{Cur01}.  To the best of our knowledge, this is the only quantum MAC (QMAC) scheme in the literature, which claims an advantage over the Wegman-Carter scheme. 
The first quantum  digital-signature scheme has been proposed by Gottesman and Chuang \cite{Got01}, and relies on the notion of a quantum one-way function, which maps classical bit strings (private keys) on quantum public keys (see also \cite{Andersson-etal06,NikPRA08,IoaMos14} for other applications of quantum public keys).  
Various authors improved on the original scheme of Gottesman and Chuang, by removing the need for quantum memory, for authenticated quantum channels, etc  \cite{Andersson-etal06,DunjkoQDS14,WallQDS15,YinQDS16,AmiriQDS16}. These develpments go beyond the scope of the present wotk, and the interested reader may refer to Refs. \cite{AA15,Pir-etal19}, for an extensive list of publications related to quantum digital signatures. It is worth noting, however, that until now all of these quantum digital-signature schemes are far less efficient than classical schemes which renders them impractical \cite{Amiri-etal18,Wang15}.  

By contrast to the thorough investigation of quantum  digital-signature  schemes, QMACs remain largely unexplored. The main question we address in the present work is whether prepare-and-measure QMACs can outperform classical schemes for 
information-theoretically secure data origin authentication.
To address this question we define a rather general theoretical 
framework for symmetric prepare-and-measure QMACs, which is absent from the 
literature. Subsequently, we show that  such schemes cannot do better 
than their classical counterparts. This result is rather general, and implies 
that the QMAC scheme of Curty and Santos cannot outperform classical 
MACs.  

The paper is organized as follows. For the sake of completeness, Sec. \ref{sec2} contains a brief summary of unconditionally secure classical MACs. In Sec. \ref{sec3} we discuss  prepare-and-measure QMACs, and our main results are presented in Sec. \ref{sec4}. 
A summary with concluding remarks is given in Sec. \ref{sec5}.

\section{Unconditionally secure classical MACs} 
\label{sec2} 

Two honest  users, Alice and Bob, share a common secret  random key $k$, which is uniformly distributed over the set ${\mathbb K}$.  The distribution of the secret key is not of our concern here, since it can be achieved by various classical or quantum means that  go beyond the scope of data origin authentication.  

Alice wants to send an authenticated message $m$ to Bob, and let ${\mathbb M}$ denote the message space, with $|{\mathbb M}|>1$.  To authenticate the message, she evaluates the tag 
$t$, through a publicly known function  $t := h(k,m)$.  
The message  and the tag $(m,t)$ are sent to Bob over a classical channel \cite{remark1}. 
In general, as a result of forgery or noise, Bob receives $(m^\prime,t^\prime)$ and he accepts the message only if $t^\prime = h(k,m^\prime)$.

In the framework of MACs, we can define the deception probability $P_l^{({\rm D})}$, to be the 
probability for an adversary (Eve) to produce a successful forgery after observing 
$l$ valid message-tag pairs. In particular, we can distinguish between an {\em impersonation} and 
a {\em substitution} attack (also known as key-only and chosen-message attacks, respectively). 
In the former attack, the adversary does not have access  to any message-tag pair (i.e., $l=0$), and 
her task is to create a pair, without knowing the secret key $k$,  that will pass Bob's verification. 
In the substitution attack, Eve has access to a single valid message-tag pair $(m,t)$, i.e. $l=1$, and 
her task is to produce another message $m^\prime\neq m$, such that $t^\prime = h(k,m^\prime)$. 
It is straightforward to generalize the substitution attack, by giving Eve access to $l>1$ valid 
message-tag pairs.

\begin{definition}
A MAC is 
{\em $l$-time ${\varepsilon}$-secure}, if for all (including unbounded) adversaries, $P_l^{({\rm D})}\leq \varepsilon$.
\label{definition1}
\end{definition}
Note that there is no MAC scheme with $P_0^{({\rm D})}=0$. 
Although Eve does not know the actual secret key, she may  guess the correct tag for her message 
$m$ with probability at least $1/|\tilde{\mathbb T}|$, 
where $\tilde{\mathbb T}$ denotes the set of different possible tags and it is assumed to be the same for all messages \cite{book3,book4,Abidin2013}.  Hence, we have 
\bea
P_0^{({\rm D})}\geq \frac{1}{|\tilde{\mathbb T}|},
\eea
and the best we can expect from a MAC scheme is to be 
$l$-time $1/|\tilde{\mathbb T}|$-secure. 
One can readily show the following fundamental theorem \cite{book4,WC81}

\begin{theorem} 
\label{theorem1}
An $l$-time $\varepsilon$-secure MAC must have keys of length at least $(l+1)|\log_2(\varepsilon)|$. 
\end{theorem}

A $1$-time $1/|\tilde{\mathbb T}|$-secure MAC can be obtained by means 
of the Wegman-Carter construction, and the use of a strongly universal hash 
function $h_{k}$ \cite{book3,book4,WC81}, which is chosen at random from a class of such functions, 
and it is identified uniquely by the shared secret key $k$.  In this case we have   
\begin{subequations}
\bea
P_0^{({\rm D})} = P_1^{({\rm D})} = \frac{1}{|\tilde{\mathbb T}|},
\label{DeceptionClassical_su:eq}
\eea
which are the lowest deception probabilities one can aim at \cite{remark2}. 
However, the cost one has to 
pay for attaining these probabilities is that the key scales linearly with the length of the message. 
More efficient 1-time $\varepsilon-$secure MACs can be constructed by means of $\varepsilon-$almost 
strongly universal hash functions, for some $\varepsilon> 1/|\tilde{\mathbb T}|$. For instance, in this context 
Wegman and Carter have shown that one can have a 1-time $2/|\tilde{\mathbb T}|-$secure MAC with a key  length which scales linearly with the length of the tag, and logarithmically with the length of the 
message i.e.,  $|k|\simeq 4 \log_2(|\tilde{\mathbb T}|) \log_2[\log_2({\mathbb M})]$.  
Various $\varepsilon-$almost strongly universal hash functions have been proposed in the literature 
(e.g., see references in \cite{book3,book4,Abidin2013}), 
and one can readily build a MAC scheme based on each one of them, with 
\bea
P_0^{({\rm D})} = \frac{1}{|\tilde{\mathbb T}|}, \quad P_1^{({\rm D})} = \varepsilon. 
\label{DeceptionClassical_asu:eq}
\eea 
The precise value of the security parameter $\varepsilon$ and the length of the key vary from scheme to scheme. Strongly universal functions can be viewed as $1/|\tilde{\mathbb T}|-$almost 
strongly universal hash functions, and it is only in this case that $\varepsilon = 1/{|\tilde{\mathbb T}|}$.
Finally, universal hash functions have been also exploited in the development of 
$l-$time unconditionally secure MACs, with $l>1$. 
\end{subequations}

The main question we address in the following section is whether 
there are  QMACs which can, in principle, compete with $\varepsilon-$secure classical MACs, in terms 
of the achievable deception probabilities and/or perhaps the required key lengths. 
A corollary of  theorem \ref{theorem1} is that 
there is no MAC with unbounded-length keys that can provide  
information-theoretic security for an unbounded number of 
messages. Hence, following standard textbooks in the 
field \cite{book3,book4} 
as well as existing work on QMACs \cite{Cur01} and 
quantum  digital-signature schemes 
\cite{Andersson-etal06,DunjkoQDS14,WallQDS15,YinQDS16,AmiriQDS16,AA15,Pir-etal19},
we will focus on the most basic scenario pertaining to the authentication of a single message. 
Moreover, although losses and imperfections during transmission are inevitable in the realization of any quantum protocol, 
it is always of vital importance to know beforehand if a task can be achieved at least ideally, in the most basic communication scenario. In this spirit, we address the aforementioned question for an ideal 
scenario i.e.,  in the absence of noise and imperfections during the transmission of the quantum states. 


\section{Theoretical framework for unconditionally secure prepare-and-measure QMACs}
\label{sec3}
We begin with a rather general theoretical framework for 
unconditionally secure prepare-and-measure QMACs, which is absent from the literature. 
 For the sake of consistency, the adopted framework follows closely the one for classical MACs that was summarized in the previous section.   

Two honest  users, Alice and Bob, share a random secret key 
$k$, uniformly distributed over the set ${\mathbb K}$, and Alice wants to send an 
authenticated classical message $m\in {\mathbb M}$ to Bob, where $|{\mathbb M}|\geq 2$.  
The key is  independent of the message, and we have the condition 
\bea
{\rm Pr}(k|m) = {\rm Pr}(k)=\frac{1}{|\mathbb K|},\,\, 
\forall\, m\in{\mathbb M}.
\label{cond1:eq}
\eea

We will assume that the tagging of the message involves the state of a $d-$dimensional 
quantum system, which is initially prepared in some publicly known quantum state 
$\ket{\Psi_{\rm in}}$.  Typically, a QMAC involves a publicly known set of 
(unitary) tagging operations $\{\hat{\cal E}_\tau~:~\tau\in {\mathbb T}\}$, where the classical 
label is a function of the key and the message i.e., $\tau = f(k,m)$ with publicly known 
$f~:~{\mathbb K}\times {\mathbb M}\to {\mathbb T}$. 
For a given pair $(k,m)$, the  operation $\hat{\cal E}_\tau$ is applied on initial state of the system thereby obtaining the quantum state 
\bea
\ket{\Psi_\tau}:=\hat{\cal E}_{\tau}\ket{\Psi_{\rm in}}.
\label{tag_operation2:eq}
\eea 
Let ${\mathbb Q}_m:=\{\ket{\Psi_\tau}~:~\tau=f(k,m),~k\in{\mathbb K}\}$ denote the set of   possible quantum states corresponding to a given message $m\in{\mathbb M}$. 
Each quantum state is  identified  by the label $\tau = f(k,m)$, and the tagging operation 
is chosen such that for {\em any} possible message $m\in{\mathbb M}$  
there  exist distinct labels $\tau,\tau^\prime\in {\mathbb T}$ such that 
$\lambda_{\tau,\tau^\prime}:=|\langle{\Psi_{\tau^\prime}}|\Psi_\tau\rangle|>0$.  
Given that $m$ is fixed, the distinct labels correspond to distinct keys.  

The quantum state $\ket{\Psi_\tau}$ serves as a quantum tag of the message  
$m$ when Alice and Bob share the secret key $k$, 
and it is sent together with the message to Bob \cite{remark1}. 
As a result of noise or Eve's intervention, Bob will receive the pair $(m^\prime,\ket{\varphi})$, and his task is to infer whether the 
received message has originated from Alice or not. In prepare-and-measure QMACs this decision is made based solely on a local measurement, perhaps after a local operation, on the received quantum tag. That is, there is no need for additional communication between Alice and Bob, and the latter  accepts the message as authentic only if the outcome of the measurement  
is consistent with the expected quantum tag $\ket{\Psi_\tau}$.  
We are interested in protocols with deterministic decision making, where Bob always accepts (with probability 1) the message-tag pair of honest Alice, with whom he shares the secret key. 
Completeness/Correctness of the protocol implies that the same 
must also hold  if Eve succeeds in guessing a valid message-tag pair. This is because, as far as Bob 
is concerned, the only feature that  discriminates Alice from Eve, is the secret key that he shares 
with the Alice. If Eve happens to guess correctly a valid  message-tag pair, this feature ceases to exist.

In general, different keys may yield the same label (and thus quantum tag) for the same message. Hence, for a given message $m$, the key space  ${\mathbb K}$ can be partitioned into smaller non-overlapping subspaces 
\[{\mathbb K}_{m,\tau}:=\{k\in{\mathbb K}~:~\tau=f(k,m)\, {\rm for}\, m\in{\mathbb M},\,\tau\in{\mathbb T}\}.\]  
Each subspace contains all the keys that lead to the same label $\tau\in{\mathbb T}$ for the given message $m\in{\mathbb M}$, while different subspaces correspond to different labels, and thus we have as many subspaces as labels.

The following discussion  focuses on prepare-and-measure QMACs and thus from now on 
the term prepare-and-measure will be omitted for the sake brevity.  
For direct comparison to standard $\varepsilon-$secure classical MACs (see Sec. \ref{sec2}), throughout this work we consider symmetric QMACs, which treat equally all of the possible messages and keys.  Hence, $|{\mathbb K}_{m,\tau}|=L$  for all $m\in{\mathbb M}$,  and $\tau\in{\mathbb T}$, while the number of 
 labels is related to the size of the key space as follows $|{\mathbb T}| = |{\mathbb K}|/L$. 
The probability for a quantum tag $\ket{\phi}$ to be equal to $\ket{\Psi_{\tau}}$ for a given 
${m}$, is given by 
\bea
{\rm Pr}(\ket{\phi}=\ket{\Psi_{\tau}}|m)
= \frac{1}{|{\mathbb T}|}.
\label{cond4:eq}
\eea
The number of different possible quantum tags is the same for any given message $m\in{\mathbb M}$, and moreover all of the quantum tags are equally probable. There are no preferable (i.e., more probable) messages, keys or tags. Of course, the partition of the key space varies from message to message, in the sense that a key has to yield a different tag for different messages [i.e., $h(k,m)\neq h(k,m^\prime)$], and thus it will appear at different subspaces \cite{remark3}.

The impersonation and the substitution attacks in the quantum setting can be defined in complete analogy to the classical setting. 

\begin{definition}
\label{definition2}
{\bf \em Impersonation attack}.---
Eve wants to  impersonate Alice without knowing the actual secret key $k$, and without having access to any valid message-tag pair. To this end, she chooses  a pair $(m, \ket{\Psi_{\tau^\prime}})$ with 
$\tau^\prime=f(k^\prime,m)$, and sends it to Bob, hoping that Bob will accept it as a valid message originated  from Alice. 
\end{definition}

 \begin{definition}
\label{definition3}
{\bf \em Substitution attack}.---
Eve does not know the secret key ${k}$ of Alice and Bob, but she has  
access to a single valid message-tag pair $({m}, \ket{\Psi_{\tau}})$. Her task is to produce another 
message ${m}^\prime \neq {m}$, such that 
the pair $({m}^\prime, \ket{\Psi_{\tilde{\tau}}})$ with  $\tilde{\tau}=f(k,m^\prime)$ will be 
accepted by Bob as a valid message originated from Alice. 
\end{definition}

In closing, it is worth noting that the theoretical framework we have just presented includes  
the unconditionally secure classical MACs discussed in Sec. \ref{sec2}. More precisely, one recovers 
the classical MACs when for any $m\in {\mathbb M}$, the possible classical tags 
$t\in\tilde{\mathbb T}$ are encoded on {\em mutually orthogonal} quantum states 
i.e., $t\to \ket{\Psi_t}$ with 
$\lambda_{t,t^\prime}=|\olap{\Psi_t}{\Psi_{t^\prime}}|=0$, 
$\forall t,t^\prime\in\tilde{\mathbb T}$ with $t\neq t^\prime$. In this case, the security 
is ensured by the classical tagging, and not by the encoding on the mutually 
orthogonal states. 

\section{Results} 
\label{sec4}

At this stage we have defined a rather general theoretical framework for unconditionally-secure  QMACs. We can prove the following theorem.  

\begin{theorem}
\label{theorem2}
For any QMAC that falls within the aforementioned framework and it involves deterministic decision making, the deception probability for the impersonation attack is higher than what can be achieved by unconditionally-secure classical MACs. 
\end{theorem}

\begin{proof}
As discussed in Sec. \ref{sec2}, for unconditionally-secure classical MACs 
$P_0^{(\rm D)} = 1/|{\mathbb T}|$. We will show that this is not possible for the QMACs discussed in Sec. \ref{sec3}, when Bob always accepts (with probability 1) Alice's message. 

In view of conditions (\ref{cond1:eq}) and (\ref{cond4:eq}), all of the keys are equally probable, and for any possible message, there are $|{\mathbb T}|$ different possible quantum tags. 
For any message there  exist at least two distinct labels, say $\tau$ and $\tau^\prime$, 
such that $\lambda_{\tau,\tau^\prime} > 0$, and in this case 
Bob cannot discriminate perfectly between the quantum states 
$\ket{\Psi_{\tau^\prime}}$ and $\ket{\Psi_{\tau}}$. 

The probability for Eve to choose the valid  message-tag pair $({m}, \ket{\Psi_{\tau}})$  is 
$|{\mathbb T}|^{-1}$, and in this case Bob will certainly accept the pair. 
By contrast to the classical setting, however, in the case of QMACs 
there is also a non-vanishing probability for Bob to accept the  
wrong pair $({m}, \ket{\Psi_{\tau^\prime}})$ as valid, because 
$\lambda_{\tau,\tau^\prime}>0$.  
The maximum probability for successful cheating is given by 
\bea
P_0^{\rm (D)} 
=
\frac{1}{|{\mathbb T}|} +
\left ( 1-\frac{1}{|{\mathbb T}|}\right ) 
\max_{\tau,{\tau}^\prime,m}\{Q({\rm acc}|\tau,{\tau}^\prime,m)\} 
\label{ImpersonationProb_Q:eq}
\eea
where $Q({\rm acc}|{\tau,{\tau}^\prime,m})$ with ${\tau}^\prime\neq \tau$, 
is the conditional probability for Bob to accept 
Eve's pair, given that Eve has sent $({m},\ket{\Psi_{\tau^\prime}})$ whereas the unknown secret key shared with Alice is $k$ so that Bob expects  $\ket{\Psi_{\tau}}$ for $\tau=h(k,m)$.  

In classical unconditionally secure classical MACs, the second term in Eq. (\ref{ImpersonationProb_Q:eq}) vanishes, because Bob can perfectly discriminate between different classical tags. So, the probability for Bob to accept the wrong message-tag pair as valid is zero, thereby obtaining $P_0^{(\rm D)}=1/|{\mathbb T}|$.  
However, for the QMACs of Sec. \ref{sec3} there exists at least one message 
such that $\lambda_{\tau,\tau^\prime}> 0$. This implies a non-vanishing probability for Bob to accept the wrong pair  (i.e., $\max_{\tau,\tau^\prime,{ m}}\{Q({\rm acc}|\tau,{\tau}^\prime,{m})\}>0$), and thus $P_0^{(\rm D)}>1/|{\mathbb T}|$, which is worse than what can be achieved in terms of classical unconditionally secure MACs.
\end{proof}

Based on the above, we conclude that one can have $P_0^{(\rm D)}=1/|{\mathbb T}|$ only when  the quantum tags with distinct labels $\tau,\tau^\prime\in{\mathbb T}$ are mutually orthogonal 
for {\em any} possible message. In this case, however, we essentially deal with a classical MAC, because one can perfectly discriminate between the different quantum tags, and 
thus the security of the protocol may rely only on the choice of the classical function $f(\cdot)$. 
The same conclusion can be reached if one works with the average rather than the maximum deception probability. 
It is also worth emphasizing that these observations and arguments do not depend on the 
type or  the size of the key. This point becomes clearer in the following subsection.


\subsection{A QMAC with quantum key} 
\label{sec4A}

Curty and Santos have proposed a QMAC for the authentication of a binary message, by means of a quantum key \cite{Cur01}. 
In their protocol, Alice (A) and Bob (B) share a pure entangled state 
\bea
\ket{\Psi}_{\rm AB} = \frac{1}{\sqrt{2}}\left (
\ket{01}_{\rm AB}-\ket{10}_{\rm AB}
\right ),
\eea
which plays the role of a secret key. The message and the tag are carried by a four-dimensional quantum system ``C" (e.g. two qubits), and let  $\{\ket{\phi_j}:j=0,1,2,3\}$ be an orthonormal basis.  

When Alice wants the send message $m\in\{0,1\}:={\mathbb M}$, she sends state $\ket{\phi_m}$ to Bob, after applying a publicly known tagging operation $\hat{\cal E}$ controlled by the state of her half of the entangled state. More precisely, the tagging operation is 
given by 
$
\hat{\cal E}_{\rm AC} = (\ket{0}\bra{0}_{\rm A} \hat{\unity}_{\rm C} +
\ket{1}\bra{1}_{\rm A} \hat{\cal U}_{\rm C}), 
$ 
for a publicly known unitary $\hat{\cal U}$, and the overall state becomes
\bea
\ket{\Psi}_{\rm ABC} = \frac{1}{\sqrt{2}}\left [
\ket{01}_{\rm AB}\ket{\phi_m}_{\rm C}-\ket{10}_{\rm AB}
\hat{\cal U}_{\rm C}\ket{\phi_m}_{\rm C}
\right ].
\eea
Tracing out the entangled state, the state of the transmitted message-tag pair reads
\bea
\tilde{\rho}_m = \frac{1}{2}\left (
\ket{\phi_m}\bra{\phi_m}
+\hat{\cal U}_{\rm C}\ket{\phi_m}\bra{\phi_m}\hat{\cal U}_{\rm C}^\dag
\right ). 
\eea  

Upon receipt of the message-tag pair, Bob performs the decoding operation 
$
\hat{\cal D}_{\rm BC} = (\ket{0}\bra{0}_{\rm B}  \hat{\cal U}_{\rm C}^\dag+
\ket{1}\bra{1}_{\rm B} \hat{\unity}_{\rm C}),
$ 
and subsequently a projective measurement on the orthonormal basis $\{\ket{\phi_j}:j=0,1,2,3\}$. 
He accepts the pair if the measurement returns one of the first two elements in the basis, i.e., $j\in\{0,1\}$, and rejects it otherwise. One can readily confirm that Bob always accepts Alice's message with probability 1.

In this  protocol Alice and Bob essentially share a secret key $k\in{\mathbb K}:=\{0,1\}$ 
in the form of an entangled state. For message $m$ Alice sends to Bob $\ket{\phi_{m}}$ if 
$k=0$, and $\hat{\cal U}\ket{\phi_{m}}$ if $k=1$. The basis state $\ket{\phi_{m}}$ can be viewed as if obtained from some publicly known initial state $\ket{\Psi_{\rm in}}$ by applying 
$\hat{\cal V}_{m}\ket{\Psi_{\rm in}}:=\ket{\phi_{m}}$. Hence, the present scheme can be 
viewed as a special case of the quantum tagging  defined 
in Eq. (\ref{tag_operation2:eq}), where the function $f(\cdot)$ is the identity function,  and 
$\hat{\cal E}_{k,m} = \hat{\cal U}_{k}\otimes \hat{\cal V}_{m}$ with $\hat{\cal U}_{0} = \hat{\unity}$ and $\hat{\cal U}_{1} = \hat{\cal U}$. The label $\tau$ is essentially $\tau=(k,m)$ and for a given 
$m\in\{0,1\}$ it  can take two possible values depending on the value of $k$, 
namely $\{(0,m),(1,m)\}$. 
According to Theorem \ref{theorem2} and the related proof, the present scheme cannot achieve $P_0^{(\rm D)}=1/|{\mathbb T}|$, unless $\lambda_{\tau,\tau{^\prime}} = 0$ for all possible messages 
in ${\mathbb M}$ and $\tau,\tau^\prime\in{\mathbb T}$, with  $\tau\neq\tau^\prime$. 
This condition reduces to 
\bea
\bra{\phi_m} \hat{\cal U}\ket{\phi_m} = 0,\,\forall m\in\{0,1\}.
\label{cond_attack1}
\eea
See also appendix \ref{app1} for an example of the impersonation attack. 

Let us consider now the substitution attack. 
Suppose that Eve has access to a valid message-tag pair, which is 
sent from Alice to Bob. Eve's task is to decide whether the transmitted system is in state 
$\ket{\phi_{m}}$ or $\hat{\cal U}\ket{\phi_{m}}$. If she succeeds, then 
she knows whether Bob will apply $\hat{\cal U}^\dag$ or not. 
Given that $\hat{\cal U}$ is a publicly known unitary, in this case 
Eve can cheat successfully by changing $m$ to $m\oplus 1$ and by sending 
$\ket{\phi_{m\oplus 1}}$ or $\hat{\cal U}\ket{\phi_{m\oplus 1}}$ to Bob. In order to prevent such an attack one needs 
\be
\label{cond_attack2}
\bra{\phi_{m}} \hat{\cal U}\ket{\phi_{m}}\neq 0,\, \forall m\in\{0,1\},
\ee
so that Eve cannot unambiguously distinguish between 
$\ket{\phi_{m}}$ and $\hat{\cal U}\ket{\phi_{m}}$. However, this requirement contradicts the conditions (\ref{cond_attack1}), required for attaining $P_{0}^{\rm (D)} =1/{\mathbb T}$. 
Thus there is no unitary operation $\hat{\cal U}$ that can satisfy both of these conditions  simultaneously. 


\subsection{Decision making based on a symmetry test}
\label{sec4B}

We have seen that a broad class of symmetric QMACs 
cannot attain $P_0^{\rm (D)} = 1/{|{\mathbb T}|} $, irrespective of the 
length of the key. The natural question arises as of whether QMACs in the particular class allow  
for shorter keys than their classical counterparts, at the cost of accepting slightly larger  deception probabilities for the impersonation attack. In this context one may, for instance, accept 
$P_0^{\rm (D)} = 1/{|{\mathbb T}|}+\delta$ in Eq. (\ref{ImpersonationProb_Q:eq}), 
for some small correction $0<\delta\leq 1/|{\mathbb T}|$. 
The precise value of $\delta$ is determined by the set of overlaps 
$\{\lambda_{\tau,\tau^\prime}:\tau,\tau^\prime\in {\mathbb T}\textrm{ with } \tau\neq\tau^\prime\}$,  
as well as by the 
strategy based on which Bob accepts/rejects a message-tag pair. 
In general, $Q({\rm acc}|\tau,{\tau}^\prime,m)$ increases with increasing 
$\lambda_{\tau,\tau^\prime}$, which in turn implies increasing deviations 
of $P_0^{\rm (D)} $ from the lowest possible 
value $1/|{\mathbb T}|$. To keep $P_0^{\rm (D)} $ close to $1/|{\mathbb T}|$ one would like to have the maximum overlap $\lambda_{\max}:=\max_{k,k^\prime,m}\{\lambda_{\tau,\tau^\prime}\}$ as small as possible (close to zero) which, however, facilitates Eve's cheating by means of a substitution attack (states with different $\tau$ become nearly orthogonal).  

One way to circumvent this stumbling block for any given $0<\lambda_{\max}<1$ is to use 
message-tag pairs of the form $(m,\ket{\Psi_{\tau}}^{\otimes n-1})$. 
Hence, the probability for Bob to accept an invalid message-tag pair 
can be reduced to any $0<\epsilon \leq 1/(|{\mathbb T}|-1)$ for a suitable 
$n= {\cal O}(|\log_2(\epsilon)|)$ \cite{Got01,NikPRA08}, without the necessity for choosing $\lambda_{\max}$ close to 0.  

It is worth noting here that Bob's  task is not to  infer the precise state of the quantum tag, 
but rather to decide on whether this state equals the expected state. In other words, given $(n-1)$ 
copies of $\ket{\Psi_{\tau^\prime}}$ (received tag) and one copy of the expected tag 
$\ket{\Psi_{\tau}}$ (which can be prepared locally), Bob has to decide whether 
$\ket{\Psi_{\tau^\prime}} = \ket{\Psi_{\tau}}$ 
or $|\langle \Psi_{\tau^\prime} \ket{\Psi_{\tau}}|<1$. In the former(latter) case he 
concludes that the received message-tag pair has (not) originated from Alice, and he  accepts(rejects) it.
This task can be performed, for instance, 
by means of a symmetry-test \cite{NikIoaPRA09}, 
which is optimal (with respect to the one-sided error requirement). 
Bob never rejects a valid message-tag pair, while for $\tau\neq\tau\prime$ the maximum 
error probability is given by \cite{NikIoaPRA09,ChaPRA18}
\bea
\max_{\tau,\tau^\prime,m}\{Q({\rm acc}|\tau,\tau^\prime,{m})\} = \frac{1}{n}\left [1+(n-1)\lambda_{\max}^2 \right ]. 
\eea    
 Inserting the resulting expression into Eq. (\ref{ImpersonationProb_Q:eq}),  and asking for 
$P_0^{\rm (D)} = 1/|{\mathbb T}|+\delta$  we obtain  
\bea
n= \frac{(|{\mathbb T}| -1)(1-\lambda_{\max}^2)}{\delta|{\mathbb T}|-(|{\mathbb T}|-1)\lambda_{\max}^2},
\eea
which is meaningful for $\lambda_{\max}<\sqrt{\delta|{\mathbb T}|/(|{\mathbb T}|-1)}$.
Recalling that $\delta\leq 1/|{\mathbb T}|$ we have 
\bea
n >|{\mathbb T}|-2,
\label{n_bound:eq}
\eea
which shows that  for any chosen $0<\delta\leq 1/|{\mathbb T}|$, the 
number of copies $n$ required for attaining 
$P_0^{\rm (D)}=1/|{\mathbb T}|+\delta$ is more than $|{\mathbb T}|-2$. 

In order to have $1$-time $\varepsilon-$secure QMAC, the entropy of the tags should be at least 
$|\log_2(\varepsilon)|+I_{\rm gain}$, where $I_{\rm gain}$ is the information that Eve can 
extract from $(n-1)$ copies of the $d-$dimensional quantum state 
$\ket{\Psi_{\tau}}$, while according to  the Holevo bound, $ I_{\rm gain}\leq (n-1)\log_2(d)$.  
Given that the message is known, this entropy can originate only from 
the key, and thus we ask for $\log_2(|{\mathbb K}|) \geq |\log_2(\varepsilon)|+(n-1)\log_2(d)$.  
Using inequality (\ref{n_bound:eq}) we find that $\log_2(|{\mathbb K}|) >|\log_2(\varepsilon)| + (|{\mathbb T}|-3)\log_2(d)$, which scales linearly with $|{\mathbb T}|$. 
This is in striking contrast to known $1-$time 
$\varepsilon-$secure classical MACs with $\varepsilon \sim 1/|{\mathbb T}|$, which attain similar deception probabilities for $\log_2(|{\mathbb K}|) \sim\log_2(|{\mathbb T}|)$ (see Sec. \ref{sec2} and related references).


\section{Summary}
\label{sec5}

We have discussed unconditionally secure data origin authentication 
by means of classical and quantum resources. 
The main question we have addressed is whether, prepare-and-measure QMACs can outperform classical 
unconditionally secure MACs. This fundamental question is of pivotal importance for the field, and to the best of our knowledge, it has not been addressed adequately in the literature so far. Although losses and imperfections are inevitable in the realization of any quantum protocol, it is always  important to know beforehand if a task can be achieved at least ideally, in the most basic communication scenario. 
Hence, the above question has been addressed in the framework of 1-time authentication under 
an ideal scenario. We showed that even under such favorable conditions, a broad class of 
prepare-and-measure QMACs cannot do better than their classical counterparts, in the sense that they 
cannot attain deception probabilities $P_{0}^{\rm (D)}=1/|{\mathbb T}|$. 
This result contradicts certain conjectures that have been made in previous related work \cite{Cur01}.  
Moreover, we showed that the key length required for attaining both 
$P_{0}^{\rm (D)}=1/|{\mathbb T}|+\delta$ (with $0<\delta\leq 1/|{\mathbb T}|$) and   
$P_{1}^{\rm (D)}\sim 1/|{\mathbb T}|$, increases linearly with the size of the tag space 
$|{\mathbb T}|$, as opposed to the logarithmic scaling in known classical schemes with similar deception probabilities.

The present analysis  and results pertain to a particular, yet rather broad class of symmetric prepare-and-measure QMACs. 
From practical point of view, such a type of QMACs is of particular interest, because the receiver can decide 
on the authenticity of the  message based solely on the received message-tag pair, and there is no need for  
additional communication between  the sender and the receiver. 
The present work may serve as a benchmark for future work 
in the field. More precisely, the generality of the results  suggests  that any 
future  efforts for the development of unconditionally secure QMACs, which 
can outperform their classical counterparts, should focus on protocols 
outside the present class of protocols (this is for instance the case of interactive and/or entanglement-assisted QMACs).  An alternative possible direction of research involves the use of physical unclonable functions with quantum readout (e.g., see \cite{puf1,puf2} and references therein).
	
\section*{Acknowledgements}
Funded by the Deutsche Forschungsgemeinschaft (DFG, German Research Foundation) – SFB 1119 – 236615297.

\appendix

\section{Security of the Curty and Santos QMAC}
\label{app1}

In the absence of imperfections and cheating, the decoding operation applied by Bob
essentially undoes the encoding operation performed by Alice and leaves Alice and Bob with the  state 
\bea
\ket{\Psi}_{\rm ABC} = \frac{1}{\sqrt{2}}\left (
\ket{01}_{\rm AB}-\ket{10}_{\rm AB}
\right )\otimes\ket{\phi_m}_{\rm C}. 
\eea
We consider now  the impersonation  attack introduced in Sec. \ref{sec3}. 

Eve does not have access to a valid message-tag pair, and she sends to Bob some state $\ket{\psi}$.  As a result, Bob's decoding operation is not canceled by Alice's decoding operation, and the state shared between Alice and Bob reads, 
\bea
\ket{\Psi}_{\rm ABC} = \frac{1}{\sqrt{2}}\left [
\ket{01}_{\rm AB}\ket{\psi}_{\rm C}-\ket{10}_{\rm AB}\left ( \hat{\cal U}_{\rm C}^\dag\ket{\psi}_{\rm C}\right )
\right ], 
\eea
where $\ket{\psi}$ denotes the state sent by Eve. Hence, the reduced state of the message-tag pair on which Bob performs the measurement is 
\bea
\tilde{\rho}_{\rm C} = \frac{1}{2}\left (
\ket{\psi}\bra{\psi}
+\hat{\cal U}^\dag\ket{\psi}\bra{\psi}\hat{\cal U}
\right )_{\rm C}.
\eea 

The probability for Eve to deceive Bob is given by 
\bea
P_0^{\rm (D)}(\psi) &=& \sum_{j=0,1} \bra{\phi_j}\tilde{\rho}_{\rm C}\ket{\phi_j}\nonumber\\
&=&\frac{1}{2} \sum_{j=0,1} |\langle \phi_j \ket{\psi}|^2+
\frac{1}{2} \sum_{j=0,1} |\bra{\psi} \hat{\cal U}\ket{\phi_j}|^2,
\eea
and depends on Eve's state $\ket{\psi}$. 
If Eve sends $\ket{\varphi_{j^\prime}}$ with $j^\prime\in\{0,1\}$,
 then $P_0^{(D)}(\phi_{j^\prime})$
is larger than or equal to $1/2$, because 
$|\olap{\varphi_j}{\varphi_j}| = 1$ and $|\olap{\varphi_j}{\varphi_{j^\prime}}| = 0$ for 
$j\neq j^\prime$.  The equality is attained only if the unitary operation used in the protocol satisfies  
\bea
\bra{\phi_{j^\prime}} \hat{\cal U}\ket{\phi_0}=\bra{\phi_{j^\prime}} \hat{\cal U}\ket{\phi_1}=0,
\eea
for all $j^\prime\in\{0,1\}$. 
One immediately sees that this condition contradicts condition (\ref{cond_attack2}).


\begin{thebibliography}{99}
\bibitem{remark1}
Throughout this work we focus on data origin authentication, and hence we assume that the message is sent in the clear together with the tag. 
If confidentiality is also an issue, the message has to be encrypted.
\bibitem{remark2} 
It is worth noting, and keeping in mind, 
that the calculation of  deception probabilities for classical MACs typically pertains to the worst-case 
scenario i.e., it involves a maximization over all possible cheating strategies for Eve.
\bibitem{remark3} 
If this is not satisfied, then Eve can easily cheat on Bob by substituting $m$ with $m^\prime$,  because the details of the protocol are publicly known and the  message is sent in the clear.
\end{thebibliography}
\end{document}